# THE EVALUATION OF OPEN SOURCE SOFTWARE INNOVATIVENESS

Nordine BENKELTOUM

University of Lille Nord de France, École Centrale de Lille

Cité Scientifique - CS20048

59651 Villeneuve d'Ascq Cedex

FRANCE

nordine.benkeltoum AT centralelille.fr

**ABSTRACT**

Product innovation assessment in software sector is a timely topic. Nevertheless, research on that subject is particularly scant. As a result, there is a lack of criteria to measure software innovativeness. In a context of theoretical and practical controversy in the open source field, this article assesses open source software innovativeness. Based on almost 500 cases studies and with the collaboration of 125 experts from industry, services and research fields, it suggests an innovation typology supported by the notion of *functional added value*. It provides also an *innovation modelling framework* that combines main evaluation methodologies. By showing the shortcomings of widely used innovation metrics, this research supports a new approach of innovativeness assessment specialized in each sector.

Keywords: open source software, free software, innovation assessment, Delphi

## Important information

This article has been originally published in French. The translation of this article (dedicated to non-French speakers) is largely based on automated translators (*Google Translate and DeepL*). All errors remain my own. For citation purpose, please refer to the French edited version published in *Systèmes d'Information et Management* available here:

**https://doi.org/10.3917/sim.133.0037**

## INTRODUCTION

Innovation research has long been confined to the study companies (or manufacturers) capabilities that selling those products. The work of Von Hippel (1988) offered a new perspective to innovation research, a new approach to innovation focusing on users (Von Hippel 2005). According to some authors, the user is even the unit of analysis of a new paradigm (Baldwin and Von Hippel 2011). Thus, user capabilities have attracted considerable academic attention resulting in a large body of research on the subject (Bogers et al. 2010; Von Hippel 2010). The literature has shown that user-designed innovations were mainly product improvements (Allen 1983; Nuvolari 2004; Ulhoi 2004), adaptations to needs (Franke and von Hippel 2003; Hicks and Pachamanova 2007; Lee and Cole 2003) or answers to practical problems (Franke et al. 2010; Von Hippel 2005). Furthermore, Internet has given rise to completely autonomous user communities such as open source communities (Iivari 2010; Stuermer 2009). Since the beginning of this century, a large number of scholars have taken an interest in these communities and have conducted a significant body of research (Von Krogh and Spaeth 2007).

The appreciation of innovation in the software sector is a relatively recent topic (Heredia Alvaro and Pikkarainen 2011). On the other hand, few academic work have addressed the question of evaluating innovation in the field of open source (Rossi 2009, pp., p. 162). Nevertheless, some observations can be made. First, this field is the subject of theoretical and practical controversies. From a theoretical point of view, the results of the two main quantitative studies are completely contradictory. On the one hand, Klincewicz suggests that the majority of open source software are clones of existing products and that only 1% of these constitute radical innovations (Klincewicz 2005). On the other hand, Rossi maintains that more than a third of the software studied is new to the market and adds that "*open source solutions are more innovative than proprietary solutions.*" (Rossi 2009, pp., p. 163).

From a practical point of view, we also find this debate. The quotes below come from interviews with open source practitioners.

*"My opinion on free software is that they are copying the existing, at least for those that I use the most, I rather use it as a tool box[1]. »*

*"I think most open source applications are not generally terribly innovative, and although there are exceptions, there are so many different software fields/categories, and within each category different applications are innovative in very different ways so it's hard to really compare or rank them[2]."*

*"You must be joking. Nothing comes to mind from the 15 years I've been using open source software. [...] Open source software is at best a cumbersome kludge. I need stuff that just bloody works. I do not need to spend weeks trying to do stuff in open source that I can do in a day in proprietary stuff [3]."*

Ultimately, the innovative nature of open source software lacks clarity (The Economist 2006, pp., p. 65) and it is difficult to assert that there is a causal relationship between the efficiency, quality and value of software on the one hand, and the fact that it is open source on the other (Fuggetta 2003, pp., p. 85). In addition, research on open source has important limitations, the main ones are as follows. First of all, many academics consider that open source promotes innovation (Fuggetta 2003, pp., p. 81) by giving the impression of confusion between process and innovative product. Then, the evaluation methods used in previous work are based on procedures that lack robustness. Thus, certain work propose a self-assessment of open source software innovativeness (Klincewicz 2005) or even offer an evaluation based on the judgment of only three experts (Rossi 2009). Finally, the literature reports a lack of relevant criteria for evaluating software in general (Lippoldt and Stryszowski 2009) and open source software in particular (Rossi 2009).

---

[1]Expert 7, Developer, Google Inc., online interview with author.
[2]Expert 104, Managing Director, TshwaneDJe human language Technology , online interview with the author.
[3]Expert 120, Free software consultant and film student, Freelance, online interview with the author.

Existing work has shown that open source software is developed by a wide variety of organizations (Benkeltoum 2011a). Indeed, some software is produced by communities of users (West and O'Mahony 2005), others come from the world of public research (Lakhani and von Hippel 2003), others were initially closed products (Hamerly et al. 1999), others are maintained by consortia (Spaeth et al. 2010), service networks (Feller et al. 2008) or even hybrid communities (Capra et al. 2011, p.: 145). In this research, the vast majority of software studied was created by users. In order to homogenize the sample, software from other organizations was excluded. Therefore, all software analysed in this article is designed by *" hobbyists [4]"* (Jeppesen and Frederiksen 2006).

In this context of theoretical and practical controversy, the research question of this article is the following: *how to assess open source software innovativeness?* The latter's response is twofold: on the one hand, it proposes a typology of innovation based on the notion of functional added value (FAV); on the other hand, it offers a framework for modelling innovation (CMI) that combines qualitative and quantitative approaches dedicated to the field of information technology (IT). To do this, it is structured as follows. The next section reviews research on product innovation evaluation and highlight the specificities of software sector. It is followed directly by a description of the main advances and limits of the evaluation of open source software and deals in particular with the question of measuring innovation in this field. The next section describes the methodological approach of this study. Then, a section presents the main contributions and implications of this research. Finally, the article closes with a presentation of the limits and perspectives for research.

---

[4]These are users developing products on their spare time.

**ASSESSMENT OF PRODUCT INNOVATION AND SOFTWARE SPECIFICITIES**

*The main approaches to measuring product innovation*

*Economics of innovation indicators*

The problem of measuring innovation is as old as classical economics (Garcia and Calantone 2002, pp., p. 111). However, one of the first typologies of innovation is attributed to Schumpeter, which it breaks down into five classes: products, production methods, markets, sources of supply and new niches (OECD/Eurostat 2005, pp., p. 35). An innovation is defined as the introduction of something new or at least one significant improvement in terms of product, service, process, marketing or organization (OECD/Eurostat 2005, pp., p. 54). This novelty can then be analysed from three perspectives: the firm, the market or the world (OECD/Eurostat 2005, pp., p. 67). Innovation is a complex phenomenon playing a crucial role for businesses (OECD/Eurostat 2005, pp., p. 14) and for the entire economy.

Since the early of 1990s, initiatives have been taken to measure and compare innovation internationally. In this field, there is a reference jointly developed by OECD[5] and European Community (EC): it is the "Oslo Manual", the evolution of versions testifies to the need to adapt the indicators to the changes in the economy. However, the metrics suggested in this manual have important limitations. First, the field of application is that of businesses, thereby excluding any non-market activity (OECD/Eurostat 2005, pp., p. 20-21) of which user communities are a part. This is particularly true in the software sector, where user-generated products account for a substantial proportion of production (Lippoldt and Stryszowski 2009, pp., p. 8). Then, in terms of radical innovation, there are significant differences in companies performance depending on the sector of activity (OECD/Eurostat 2005, pp., p. 45). If part of this disparity can be attributed to a difference in dynamism between sectors; another part is necessarily attributable to the metrics used to analyse radicalness. In some sectors, this has led researchers to develop *ad hoc* indicators. There is the sector of traditional activities, where indicators linked to aesthetics have been developed (Alcaide-Marzal and Tortajada-Esparza 2007), or the software sector, the particularities of which denote the need for specific metrics (Lippoldt and Stryszowski 2009).

*Indicators in innovation management*

Alongside these indicators, innovation management research offers a set of criteria for assessing product innovation[6]. There are two dominant paradigms to assess innovation: the *incremental/radical* couple and the *sustaining/disruptive* couple. Thus, incremental and radical innovations are distinguished by the type of knowledge they contain. Radical innovations contain a high degree of new knowledge while incremental innovations rely on a low level of new knowledge (Dewar and Dutton 1986, pp., p. 1422). The notion of radically new product refers to products based on a new technology (Chandy and Tellis 1998, pp., p. 476).

Sustaining and disruptive innovations are assessed according to marketing criteria (Christensen 1997; Christensen and Raynor 2003). On the one hand, sustaining innovations consist of improving products based on market criteria. On the other hand, disruptive innovations introduce a value proposition not expected by the market (Christensen 1997;

---

[5]Organisation for Economic Co-operation and Development
[6]For a relatively comprehensive review of generic metrics for evaluating innovation refer to (Garcia and Calantone 2002).

Christensen and Overdorf 2000). The concept of disruptive innovation has been the subject of intense debate in new product development. The literature has highlighted that there are no precise criteria for determining whether a technology is disruptive or not (Danneels 2004; 2006).

To conclude, the abundance of terminologies and the debate around their definition makes their use very difficult. Furthermore, as noted by researchers *"the new product development literature has primarily used dichotomous classifications to identify the type of innovation. We think these dichotomies are too simplistic"* (Garcia and Calantone 2002, pp., p. 120).

### Assessing innovation in the software sector

Innovation in the software sector is defined as a process leading to the development of a new design, a new functionality or a new application for an existing product; a new product, service or process; improvement of a product, service or process; entering a market or creating a new market (Lippoldt and Stryszowski 2009, pp., p. 10). The characteristics of innovation in this area differ from other sectors of activity in several respects. The piece of software is unique since unlike other products, it continues to evolve after its distribution through the mechanism of updates (Codenie et al. 2011, pp., p. 7). In other words, software is constantly in process. Then, it is also intangible (Codenie et al. 2011, pp., p. 7; Lippoldt and Stryszowski 2009, pp., p. 8), this means that it is not palpable, but also and above all, that it can be replicated and distributed at low cost. Finally, it is easy to design software because the market entry costs are low. All you need is a computer and programming knowledge (Codenie et al. 2011, pp., p. 8).

The metrics used to identify the innovation of tangible products such as patents (Lippoldt and Stryszowski 2009, pp., p. 86) or brands (Rossi 2009, pp., p. 155) are not relevant for software. Moreover, there are no clear and universal metrics to measure software innovation (Lippoldt and Stryszowski 2009, pp., p. 8). Consequently, the software industry requires an adaptation of existing metrics to its singularities (Klincewicz 2005, pp., p. 7) or even the creation of specific metrics (Lippoldt and Stryszowski 2009; Rossi 2009, pp., p. 155). Several authors have attempted to propose metrics adapted to the specificities of the software by focusing on the elements making it possible to identify radical innovations. Soens suggests four classes of innovation: (1) class 1: incremental innovations, (2) class 2: market-driven radical innovations, (3) class 3: technology-driven radical innovations, ( 4) class 4: innovations based on research (Soens 2011, pp., p. 41-45).

Klincewicz offers an analysis grid based on the work of Dahlin and Behrens (2005). The latter support the idea that a radically innovative technology is successful when it is: *"new, unique and has an impact on future technologies"* (Dahlin and Behrens 2005, p.: 717). According to Klincewicz, the criterion of novelty and uniqueness are satisfied if no comparable functionality existed before the date of the product's launch. Klincewicz adds the dimension of platform (operating system) and proposes the grid reproduced below.

**Table 1: Software innovation evaluation grid ( Klincewicz , 2005)**

| | New technology | New for a platform | Existing technology |
|---|---|---|---|
| **New market** | Radical invention (rupture) | | Marketing innovation |

| **Existing market** | Technology change | Platform modification | No innovation |

Rossi proposes specific evaluation criteria classified in three dimensions: the firm, the market and the world. (1) The first dimension measures innovation for the firm. It is based on a single indicator indicating whether the software is new to the firm (indicator 1). (2) The second dimension includes two indicators and measures whether the software is innovative for the market. The first criterion analyses the ability of the piece of software to meet the needs and demands of users (indicator 2). The second criterion concerns the degree of novelty of the software from a technological point of view (Indicator 3). (3) The third dimension is the most general, it contains two criteria. The first criterion questions whether the software contains a new module (Indicator 4). The second criterion relates to the novelty of the software platform (Indicator 5) (Rossi 2009, p.: 160).

## PROGRESS AND LIMITATIONS OF THE EVALUATION OF OPEN SOURCE SOFTWARE

### *General information on the evaluation of open source software*

#### *Characteristics and controversies*

Open source technologies have become essential in information systems (IS) both in theory (Fitz-Gerald 2010; Gary et al. 2011) and in practice (CIGREF 2011; Le Texier and Versailles 2009; Lindman et al. 2011; Lisein et al. 2009). While the benefits of these technologies are numerous and well documented (Thakur 2012), other features are much more controversial (Fuggetta 2003). First, with regard to the advantages, free software has a quality source code structure (Capra et al. 2011; Von Krogh and Spaeth 2007) and the organizations providing the development guarantee rapid bug fixing (Bitzer and Schröder 2005; Paulson et al. 2004). In addition, open source guarantees transparency (Spinellis et al. 2009), interoperability as well as the technological independence of companies and countries (Benkeltoum 2011a; Benkeltoum 2011b). This observation is particularly true for developing (Chen et al. 2010) and rising powers (China, Brazil) countries.

The points of controversy are also numerous (Fuggetta 2003) and concern in particular reuse practices, modularity and innovation[7]. For some academics, open source software presents a completely satisfactory rate of code reuse internally and externally (between projects) (Haefliger et al. 2008), and for others these practices can largely be improved (Capiluppi et al. 2011). Likewise, one study demonstrates that free software does not have a more modular structure than closed software (Paulson et al. 2004) while another defends exactly the opposite thesis (MacCormack et al. 2006).

#### *Assessment approaches*

IT evaluation is always a timely topic (Benlian 2011; Lippoldt and Stryszowski 2009), open source software is no exception to the rule (Lundell et al. 2010; Miralles et al. 2006). Indeed, there is a growing theoretical corpus on the evaluation of open source software where two approaches are identifiable. On the one hand, open source software is assessed via well-known metrics in IT evaluation (Miralles et al. 2006). For example, Benlian compares open and closed office suites based on the same metrics: features, price, ease of use and technical support (Benlian 2011). From an economic point of view, this software can be evaluated using the *Total Cost of Ownership* (TCO) model (Fitzgerald 2006) which takes into consideration all the costs associated with the adoption of a technology such as: training, migration costs or even maintenance (Taibi et al. 2007). While a comparison centred on the cost of acquisition (licenses) favours open source software because most of it is distributed free of charge (Benkeltoum 2011b).

---

[7]This last point will be dealt with specifically in the next section.

On the other hand, another stream of research has proven that the appreciation of open source software requires adequate[8] and enriched metrics. For example, because the source codes of most of these software are easily accessible, the evaluation can be based on an analysis of the quality (Lundell et al. 2010; Midha and Palvia 2012) and the structure of the source code (Capra et al. 2011; MacCormack et al. 2006). The evaluation of closed software cannot be carried out without decompilation, which the licenses of the latter prohibit . Similarly, many pieces of free software are not distributed by companies. It is therefore impossible to measure responsiveness or assess the services of the editor. These elements must be measured via *ad hoc* means such as the activity of the mailing lists, the quality of the source code documentation or even via the existence of community services (Del Bianco et al. 2009; Lee et al. 2009; Spinellis et al. 2009; Taibi et al. 2007). This article is part of this stream of research and builds metrics specific to open source software.

### Assessing open source software innovation

#### Theoretical controversy

In the literature, only two studies address the issue of innovation evaluation in open source in depth. The latter offer completely contradictory results. Klincewicz rates SourceForge's 500 most active software using the self-assessment method. More specifically, Klincewicz *"focuses on the importance of the changes to the entire product, perceived by the individuals developing the project"* (Klincewicz 2005, p.: 9). For this, it uses an evaluation grid based on the notion of radical innovation adapted to software (see previous section). The results show that 87.2% of the free software studied are not innovative, 10.6% introduce platform changes, 1% radical innovations, 0.8% technology changes and 0.6% innovations. marketing. This research proves that it is easy to identify non-innovative software and platform modifications but that it is complex to identify other categories of software, in particular radical innovations.

Rossi compares the solutions offered by Italian SMEs[9] based on free software and *"proprietary"* solutions (Rossi 2009). The evaluation covers a total of 134 software: 107 are closed and 27 are open source. It shows that free software is more innovative than closed software on all the criteria (see previous section): whether from the point of view of the firm, the market or the world.

#### The limits of literature

---

[8] Two European projects have succeeded (Qualoss and QualiPSo) on the issue of measuring the quality of open source software, which testifies to the complexity of the subject. For a description of QualiPSo refer to Benkeltoum (2011a, pp., p. 123-125).

[9] Small and medium enterprises

Work on the evaluation of open source software innovation has the following limitations: the apparent confusion between process and innovative product, the weakness of the evaluation procedure and the lack of relevant criteria for evaluating software innovation.

- *Limit 1: the apparent confusion between process and innovative product*

While some scholars consider open source not to represent a real change in software development (Fitzgerald 2006, pp., p. 594), others suggest that it constitutes a new approach (Fuggetta 2003, pp., p. 81). Moreover, it seems that many researchers consider that open source products are innovative because their design process is innovative, and use the terms *open source development* and *open source innovation* synonymously. The risk is to use the word *innovation* as a fad (Klincewicz 2005, p.: 3). Thus, for some open source symbolizes in itself a *breakthrough innovation* (Rossi 2009, p.: 154), an extreme case of *open innovation* (Dahlander and Wallin 2006) or a *radical innovation* (Bonaccorsi et al. 2006, p.: 1086).

This mode of development, with geographically distributed participants, has sometimes been described as *"E-Innovation"* (Kogut and Metiu 2000; Kogut and Metiu 2001), *"innovation networks by and for users"* (Von Hippel 2002) or even the *"Private -Collective Model"* where participants use their own resources to constitute a collective good (Von Hippel and Von Krogh 2003, p.: 213). More recently, this model has been extended to hybrid communities linking a firm (Nokia), partners and volunteer developers (Stuermer et al. 2009). However, this model has an important limit since it does not propose a distinction between what is innovative and what is not. For example, Nokia offers competing companies to use its platform. The authors describe this action as the diffusion of innovation (Stuermer et al. 2009, p.: 180) whereas the phenomenon highlighted relates to the sharing of a generic technological base on which Nokia does not create value[10]. Nokia prefers to keep control over the application layers and does not share at this level. This case is typically similar to the strategy of integrators like Thales in the field of *middleware* which shares technologies on lower layers and keeps secret strategic parts (Benkeltoum 2011a). This joins the statement of a Nokia employee for whom *"our thing is not so revolutionary"* (Stuermer et al. 2009, pp., p. 182).

Many authors (Hicks and Pachamanova 2007; Lee and Cole 2003; Osterloh and Rota 2007; Von Hippel and Von Krogh 2003) consider all software modifications as an innovation. Thus, the return of modifications made by users to the code of free software is considered to be a phenomenon of user innovation propagation (Hicks and Pachamanova 2007, p.: 318). A study of Linux kernel development showed that developers had two functions: a quality assurance function and an innovation function. For innovation, developers suggest new functions or write patches (Lee and Cole 2003, p.: 637). Similarly, some researchers confuse production and innovation activities (Bogers et al. 2010; Osterloh and Rota 2007; Von Krogh et al. 2003). Finally, others defend the idea that the open source paradigm promotes innovation (Deek and McHugh 2007; Ebert 2007; Ebert 2008; Janamanchi et al. 2009; Von Krogh and Spaeth 2007).

- *Limit 2: an evaluation procedure with weak robustness*

A qualitative assessment of open source software innovation highlighted that open source software is both innovation and copying (Deek and McHugh 2007, pp., p. 7). On the one hand, free software copies closed software. On the other hand, they are the source of technology that enabled the creation of the internet (Fitz-Gerald 2010). Nevertheless, Deek and McHugh do not detail the criteria used to distinguish between innovative and non-innovative open source software.

---

[10] The authors themselves speak of *"generic frameworks"* ( Stuermer , Spaeth and von Krogh, 2009: 185).

Klincewicz's study covers 500 open source software using the self-assessment technique (Klincewicz 2005). Rossi evaluates 134 software (107 closed and 27 open source) based on the evaluation of three experts (Rossi 2009, pp., p. 159). In addition, the research question of the study is: *"Do programs based on FLOSS solutions are more innovative than proprietary [software]"* (Rossi 2009, pp., p. 153)? However, it should be emphasized that Rossi evaluates software solutions based on free software rather than free software itself. The resulting product is rather a combination of free and non-free software, i.e. a hybrid solution (Bonaccorsi et al. 2006).

- *Limit 3: the lack of criteria to assess software innovation*

The literature has highlighted the lack of relevant criteria for assessing innovation in all sectors of the economy (Le Masson et al. 2010; Rossi 2009) and in particular for software (Lippoldt and Stryszowski 2009, pp., p. 8). Similarly, Deek and McHugh do not reveal the criteria used to distinguish innovative from imitative (Deek and McHugh 2007). In his study, Klincewicz uses untested criteria to gauge the innovation of 500 software programs (Klincewicz 2005). Rossi proposes an evaluation of 27 free software based on criteria not empirically validated (Rossi 2009). In this latest research, more than a third (35%) of products are considered *"new to market"* and more generally, a third of products receive high scores for innovation (Rossi 2009, pp., p. 162) which is extremely rare for an industry. For example, the new product development literature estimates that only 10% of innovations are radical (Garcia and Calantone 2002, pp., p. 120). Moreover, the criteria of radicalness differ for users and manufacturers (Afuah 2000; Bogers et al. 2010).

# RESEARCH METHODOLOGY

## *Overview of research*

### *Positioning and objectives*

This article will be based on the definition of Lippoldt and Stryszowski who emphasize that an innovation in the field of software can be purely functional (Lippoldt and Stryszowski 2009, pp., p. 10). On the other hand, contrary to Klincewicz, this research will not consider that the proposal of a new and unique functionality constitutes a radical innovation (Klincewicz 2005) since the notion of radical innovation focuses on the use of a new technology (Chandy and Tellis 1998). For these reasons, the notion of functional innovation will be used because it is more in line with the nature of the innovation in question. The objective of this research is to provide a typology of functional innovation associated with metrics specific to open source. To do this, the notion of *added value* will be used. It designates an added value provided by software vis-à-vis existing products.

IS research has shown that the combination of qualitative and quantitative methodologies is a powerful tool (Agerfalk and Fitzgerald 2008; Mingers 2000; Mingers 2003; Pinsonneault and Kraemer 1993) and especially in open source (Feller et al. 2008; Haefliger et al. 2008; Lee and Davis 2003). This research is based on multimethodology. An approach that combines different methodologies or techniques (quantitative or qualitative) in order to take advantage of each method (Mingers 2000). It is defined as *"a set of guidelines or activities to help people carry out research or interventions"* (Mingers and Brocklesby 1997, p.: 490). Four arguments are in favour of multimethodology: (1) the complexity of reality events, (2) the fact that research is most of the time divided into stages, (3) that the use of multimethodology is common in practice and finally (4) because it constitutes a contemporary form in practice (Mingers and Brocklesby 1997, p.: 492).

Mingers and Brocklesby (1997, p.: 491) propose five levels ranging from what the authors call isolationism to multi-methodology (see Table 2).

**Table 2: from isolationism to multi-methodology**

| Principle | Description |
|---|---|
| Isolationism | Use only one method or technique from a paradigm |
| Improvement | Improve a methodology |
| Selection | Choose the appropriate methodologies according to the context |
| Combination | Combining methodologies for a purpose |
| Partition and combination | Partition and combine methodologies |

### *General research design*

The methodology of this research is broken down into four stages: (1) the qualitative definition of types of functional innovation on the basis of large-scale case studies (294 cases), interviews with experts and a review literature; (2) an evaluation of 25 open source software by 125 experts; (3) a comparison of qualitative and quantitative classifications; (4) modelling types of functional innovation. Each step is based on different methods that contribute to the overall research objective (Table 3).

**Table 3: contribution of each step to the research objectives**

| Stage | Title | Goals |
|---|---|---|
| 1 | Qualitative definition of types of functional innovation | - Field exploration<br>- Identification of a priori constructs<br>- Qualitative categorization of free software |
| 2 | Expert evaluation of open source software | - Variable validation<br>- Validation of assumptions |
| 3 | Comparison of qualitative and quantitative classifications | - Assigning a type based on expert judgment<br>- Classification validation |
| 4 | Modelling the types of functional innovation | - Modelling types of functional innovation in open source |

In this research, different methods were combined at each stage. They will be presented as this study progresses. The table below offers an overview of the principles of these methods, a brief description of the role played by them at the different stages of research.

**Table 4: combined methods for the study of functional innovation**

| Methods | Principles | Parts used and description | Steps | Literature |
|---|---|---|---|---|
| Multiple case study | Understand the dynamics between cases | - "theoretical sampling": selection of 294 open source software<br>- use of the principles of contradiction, extension and replication | 1, 2 | (Eisenhardt 1989; Eisenhardt and Graebner 2007) |
| Delphi technique | Expert evaluation | - expert review: 25 open source software<br>- statistical aggregation: 125 evaluations on 25 open source software | 2 | (Danneels 2004; Rossi 2009; Rowe and Wright 1999) |
| Grounded theory | Discovering theory from data | - qualitative coding of types of innovation | 1 | (Glaser and Strauss 1967; Sigfridsson and Sheehan 2011) |
| Statistics | Use of statistical tools | - variance analysis: validation of variables<br>- principal component analysis: comparison between qualitative classification and that of experts<br>- Discriminant analysis: testing the integrity of qualitative classification and modelling innovation | 2, 3, 4 | (Burns and Burns 2008; Capra et al. 2011; Lee et al. 2009) |
| Snowball sampling | Identification of respondents using previous respondents | - distribution of the questionnaire among the experts | 2 | (Shah 2006) |
| Netnography | *"Ethnography adapted for the study of online communities"* (Kozinets 2002) | - online discussion: interviews with developers and *key informants*<br>- study threads | 1, 2, 3 | (Kozinets 2002; 2006; Sigfridsson and Sheehan 2011) |
| Self-evaluation | Evaluation based on selected criteria | - self-assessment of 294 open source software<br>- self-assessment of 152 open source software | 1, 2 | (Chen et al. 2010; Del Bianco et al. 2009; Fearon and Philip 1998; Klincewicz 2005; Rowe and Wright 1996) |

*Detailed research steps*

*Step 1: qualitative definition of types of functional innovation*

- *Goals*

The purpose of this first step is to propose a typology of functional innovation in open source software based on a qualitative analysis.

- *Data*

Evaluation criteria were defined on the basis of a sample of 294 open source software. These software were selected largely on the basis of a list of 231 free software to be tested under the free ReactOS operating system[11]. This list was supplemented by free software used by the author or mentioned by experts. For each free software, information has been gathered about: its main functionalities, its advantages compared to existing closed software and its origin. These data were collected thanks to the exchange of more than 524 emails[12] with the founders or *key players* in the project (Agerfalk and Fitzgerald 2008, pp., p. 392) but also information available on the official site of each software (Stewart and Gosain 2006, pp., p. 299) by triangulating each time the information with other sources (Wikipedia, Freecode, forums, etc.).

In-depth interviews with open source professionals were conducted. Three experts played a particularly fundamental role in defining the types of innovation in the field of free software ( **Table 5**). 1) The first expert is an industrial researcher from Alcatel-Lucent who *"works in R&D in the IP (Internet) world"* and *"regularly files patents"*. He has more than 10 years of experience in development and is the author of several free software. Around 67 emails were exchanged with this expert. 2) The second expert is a project manager within Thales D3S. He has more than ten years of development experience. He holds positions of responsibility in various European projects dedicated to free software. He has a long experience in development, notably for a Linux distributor. About 20 e-mails were exchanged, an interview of about two hours was conducted and informal discussions conducted during participation in a symposium dedicated to open source technologies (over four days). 3) The third expert is a development engineer for an SME publishing free software. He is also a member of the board of directors of an association for the promotion and defence of free software. He has almost ten years of experience in software development. About 190 e-mails were exchanged, an interview lasting about an hour and a half and informal discussions within the framework dedicated to free software were carried out. The members of the board of directors (composed of 14 open source experts) of this association[13] have largely contributed to the improvement of this research by offering a critical analysis of it.

**Table 5: summary of the profiles of the three experts**

| ID | Function | Business | Community | Experience | Sources of information |
|----|----------|----------|-----------|------------|------------------------|
| 1 | industrial researcher | Alcatel Lucent | Author of several free software | - 10 years in development <br> - Regular filing of patents | - 67 emails |

---

[11] ReactOS is an operating system aiming to be fully compatible with 32-bit Windows applications.
[12] Calculation based on a count of electronic archives.
[13] The association has approximately fifty institutional members and 350 individual members.

| 2 | Project manager | Thales D3S | OW2 member Mandriva | - 10 years in development<br>- European project manager | - 20 Emails<br>- 2 hours' interview<br>- Informal discussions (4-day seminar) |
| 3 | Development engineer | Free Software Editor | Board member of AFUL | - 10 years in development | - 190 emails<br>- 2 hours' interview<br>- Informal Discussions |

- *Analyses*

The data collected during the qualitative phase were coded according to the following iterative process:

1) study of the license to ensure that it is compatible with the criteria of the Free Software Foundation [14]: 6 software programs were excluded;
2) verification that the software was indeed designed by users: 35 software released by companies were excluded;
3) identification of the domain and subdomain to which the software belongs based on its definition;
4) prior art search by contact with founders, developers or public data;
5) identification of the presence or absence of added value(s) introduced by the software from a functional point of view (sources identical to 4);
6) separation of software into two types: with added value (55) and without functional added value (200);
7) qualitative classification of software based on the type of added value.

For the seventh phase, software was classified using the grounded theory approach which aims to emerge theories from data (Glaser and Strauss 1967; Sigfridsson and Sheehan 2011). Thus, the data were analyzed manually by maximizing the phenomena of inversion, extension and replication (Eisenhardt and Graebner 2007, pp., p. 27). More precisely, it was a question of: distinguishing software providing added value (AV) from that providing none (reversal of phenomenon); create new types of VA when the VA provided by a product did not fall into the previously identified categories (extension of the phenomenon); identify the repetition of the type of VA (replication of the phenomenon). This classification is also based on the opinion of three experts ( **Table 5** ) and a review of the functional innovations mentioned in the literature. Ultimately, a process of self-assessment, a method where the researchers themselves define the criteria used for the assessment (Chen et al. 2010; Del Bianco et al. 2009; Fearon and Philip 1998; Klincewicz 2005; Rowe and Wright 1996), was followed. This process can be schematized as below.

**Figure 1: qualitative classification (step 1)**

---

[14]According to the FSF, software is considered free if its license confers four freedoms on the user: execution, study, modification and distribution.

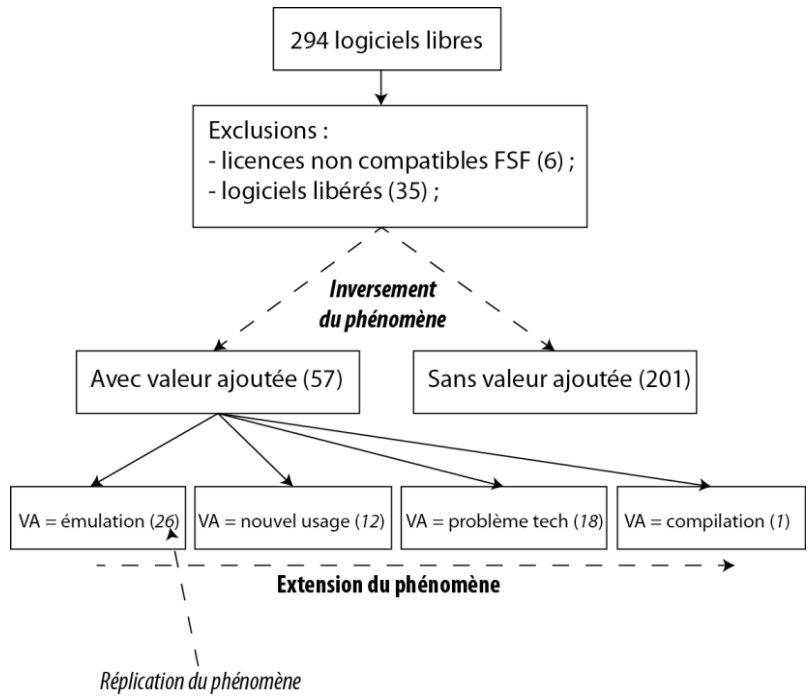

- *Results*

This process made it possible to identify a priori constructs (Eisenhardt 1989; Miles and Huberman 1994). These constructs were then used to create an evaluation questionnaire (see step 2), five types of functional innovation thus emerged from the data: (1) The free alternative, an alternative software to an existing closed software. This category gathers all the software which did not bring any added value from the functional point of view apart from the fact of being free. (2) Emulator, a *virtualized version* of hardware or a set of software applications. It is innovative since it allows users to replace equipment that is no longer manufactured because it is considered obsolete. (3) The package, software grouping together several free software in a single block, most of the time to facilitate the use, installation and/or configuration of complex software. (4) The adaptation part, software that solves a technical problem that has never been solved before. (5) New use orientation, software introducing a new use concept. This typology is based on both data and/or literature on open source. In addition, examples make it possible to illustrate this typology concretely ( Table 6).

**Table 6: categories resulting from the qualitative analysis**

| Kinds | Detailed example | Other examples | Literature |
|---|---|---|---|
| Alternative | The GNU Project is a free alternative to Unix (Stallman 1983) | Linux Kernel, Gnu Privacy Guard | (Hars and Ou 2002; Osterloh and Rota 2007; Spiller and Wichmann 2002) |
| emulator | PCSX2 simulates the operation of the Playstation 2 | Cassini, Pear PC | (Von Krogh and Spaeth 2007) |
| package | The AMP package bundles Apache, MySQL, Perl, and PHP | Debian, Ubuntu | (Deek and McHugh 2007) |
| Adapter piece | Mono allows running .Net applications on Linux/Unix systems | Wine , Samba | (Hicks and Pachamanova 2007; Stallman 2002) |
| New use oriented | Mute is P2P (Peer to Peer) software allowing its users to preserve their anonymity | Nmap , BitTorrent | |

*Step 2: Expert evaluation of open source software*

- *Goals*

The purpose of this step is to validate the relevance of the variables derived from the qualitative study. It is for this purpose that hypotheses were derived from step 1:

- *H1: some free software represents alternatives to existing closed software;*
- *H2: some free software emulates the operation of hardware or a set of software applications;*
- *H3: some free software combines a set of free software;*
- *H4: some free software solves one or more previously unresolved technical problem(s);*
- *H5: certain free software creates a new concept of use.*

All of these hypotheses will be tested using an evaluation based on the Delphi technique (Rowe and Wright 1999). This technique is used when a lack of appropriate data is observed and when human judgment is required (Nambisan et al. 1999, pp., p. 374). There are four conditions for considering a technique as "Delphi": anonymity, iteration, controlled feedback and statistical aggregation of responses. Since this research was conducted in collaboration with experts distributed globally (26 countries represented), only a variation of the Delphi technique was adopted based on the recommendations of Rowe and Wright (1999 (1999, pp.: 354-355).

In short, step 2 combines qualitative and quantitative approaches. More specifically, it involves: on the one hand, selecting software whose type has previously been identified via the qualitative process already presented and, on the other hand, recruiting experts in order to submit to them a questionnaire enabling them to evaluate previously selected software.

- *Data*

A questionnaire ( **Appendix 2**) was designed with a dual objective: on the one hand, to test the aforementioned hypotheses; on the other hand, compare the qualitative classification with that emanating from open source experts (step 3). This questionnaire was designed in collaboration with experts who commented on the different versions by email. A summary of the experts' contributions to this questionnaire is provided in **Appendix 2**.

To do this, a list of 152 free software was designed from the 100 most popular software from the SourceForge (100) and Freecode (100) sites by eliminating duplicates. The selection of popular free software is justified by the fact that the software must be sufficiently known by experts to be evaluated satisfactorily. Indeed, a pilot study was conducted, like that of Feller et al. (2008), during which five experts evaluated 25 popular and non-popular free software. However, some software was not evaluated because it was unknown to the respondents. As one expert pointed out during this phase, *"it's difficult to give a useful opinion on [software] that we don't know"* (Open Source Architect, Thales). Moreover, the literature has shown that popularity is an important parameter to assess the quality (Capra et al. 2011, p.: 146)and success of open source software (Midha and Palvia 2012). Finally, previous works have highlighted the need to analyze the innovation of the most popular open source products (Rossi 2009, pp., p. 165).

For each software, information has been gathered about: its main functionalities, its added value(s) compared to existing closed software, its origin. These data were collected through public information, supplemented by the exchange of 50 messages with key players in the projects. They were coded using the same process as step 1 (7 phases of the coding process). Except for the last phase (qualitative classification) of the data coding process which also led to the selection of software for expert evaluation. This classification is summarized below.

**Figure 2: Qualitative classification and selection (step 2)**

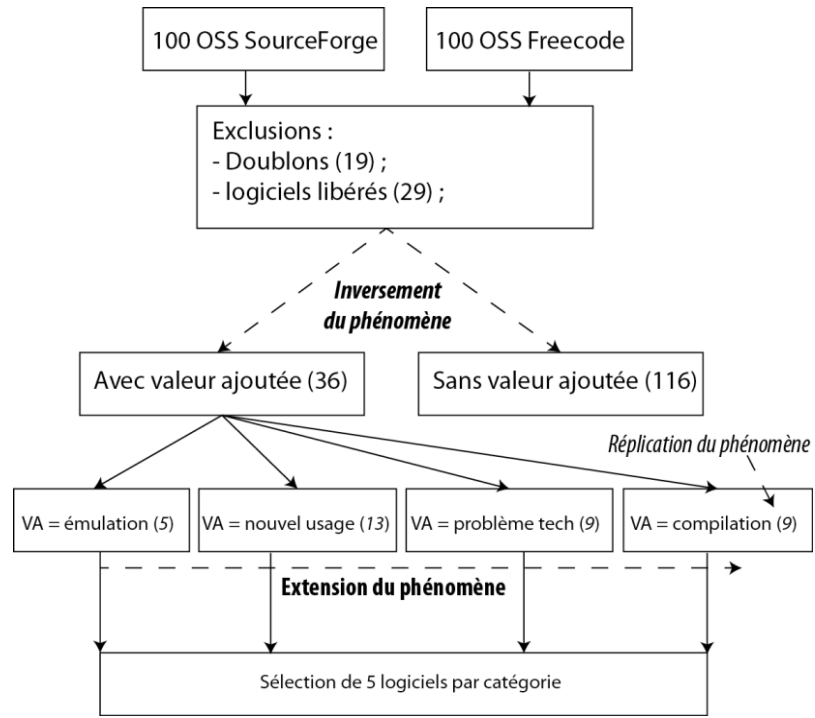

As explained in the figure above, a selection of five software from each category beforehand (classified qualitatively) was made, i.e. 25 software in total (see Table **Table 7**. This selection was based on the principles of theoretical sampling (Eisenhardt and Graebner 2007)in order to obtain a sample maximizing the phenomena of inversion (selection of software with and without added value), extension (software of each type) and replication (selection of 5 software per type).

**Table 7: Software selected for evaluation**

| Name | Definition | Qualitative classification |
|---|---|---|
| 7-Zip | File compression utility | Free alternatives |
| ALSA Driver | sound driver for linux | Adapter piece |
| BitTorrent | P2P File Sharing Protocol | New use oriented |
| DOSBox | Dos Emulator | emulator |
| FileZilla | FTP client | Free alternative |
| Gimp | Image editing program | Free alternatives |
| KNOPPIX | Linux distribution on live CD | package |
| Linux NTFS | NTFS File System Driver for Linux | Adapter piece |
| MiKTeX | Distribution of TeX | package |
| MinGW | Adaptation of GNU development tools to W32 systems | Adapter piece |
| ncurses | System V emulation | emulator |
| Nmap | security scanner | New use oriented |
| PDFCreator | PDF file creation tool | Free alternative |
| Peer Guardian | Privacy tool for P2P | New use oriented |
| PHP | Scripting language | New use oriented |
| pidgin | Multi-protocol instant messaging client | New use oriented |
| PortableApps.com | Portable Application Package | package |
| Samba | System interoperability between Linux/Unix and Windows systems | Adapter piece |
| ScummVM | Scumm System Emulator | emulator |
| Util -linux | Suite of utilities for Linux systems | package |
| VirtualDub | video editing software | Free alternative |
| VisualBoyAdvance | GameboyAdvance emulation | Emulator |
| Wine | Implementing Windows for Unix/Linux Systems | Adaptation piece |
| XAMPP | Apache, MySQL, PHP and Perl installer | package |
| ZNES | Super Nintendo emulator | Emulator |

Concretely, a questionnaire was administered to 125 open source experts (see **Appendix 2**) recruited during various interactions with the field. A good number come from organizations studied as part of a thesis. Many others were selected on the principle of *snowball sampling* , which consists of identifying new respondents through old respondents (Shah 2006). Around 450 emails [15]were exchanged with the experts. Free software user associations also served as relays by posting a link to the questionnaire made available online. Because software evaluation is a complex task, the respondents were predominantly practitioners. Ultimately, this study brings together free software specialists from various institutions: 70% work in industry, IT services (Thales, Google, Alcatel-Lucent, Motorola, etc.) or research centers (INRIA , NASA , the Danish Navy, CNRS, etc.) from all over the world (see Appendix 1). For each software, the experts expressed their opinions on five dimensions (see **Appendix 2**for the questionnaire items) using a five-item Likert scale ( *"I completely agree"* ; *"I agree »* ; *"I don't agree"* ; *"I don't agree at all"* and finally *"I don't know"* ).

- *Analyses*

---

[15]Calculation based on a count of electronic archives.

For each question, all the "I completely agree" and "I agree" were added up as many experts expressing their agreement with the proposed statement. In order to have comparable data, the evaluation data have been aggregated for each software and reported on the same scale (the percentage). To do this, the number of experts who responded to each item was divided by the number of respondents (see Appendix 4) which represents 3125 aggregated evaluations (125 expert evaluations on 25 software). In short, the answers of the experts were transformed into a percentage of agreement with the question asked. For each software, an a priori type was coded in the form of a nominal variable. After that, an analysis of variance (ANOVA) and a principal component analysis (PCA) were carried out in order to validate the hypotheses and validate the variables resulting from the qualitative phase.

- *Results*

The ANOVA shows significant differences for the means of all the dependent variables (P<0.000) or (P<0.05) (see **Annex 5**). This proves that the variables are relevant and correlated with the experts' assessments, thus validating all the starting hypotheses (H1 to H5).

The PCA calculated 5 axes explaining all of the variance ( **Annex 5**). However, by applying the *Guttman -Kaiser rule* , we only retain the first two factors since this rule provides that only the axes having an eigenvalue (or *eigenvalue* ) greater than 1 must be retained (Rietveld and Van Hout 1993, p. p. 273). Likewise, the elbow test (or *scree plot test* ) shows that only the first two factors have an eigenvalue greater than 1 ( Appendix 6). Consequently, the two factors retained (see Table 8) explain respectively 0.391 and 0.339 of the variance. Combined, they represent almost 0.73 of the variance, which is within the acceptable average (Rietveld and Van Hout 1993, p. p. 273). Because the first component focuses on the creation of a new function and the resolution of a technical problem, it was called the degree of functional innovation. On the same principle, the second component focusing on the reproduction of existing products was called degree of functional reproduction ( **Figure 3**).

In addition, we also see two major negative correlations. The *"free alternative"* variable is negatively correlated with component 1. This phenomenon is consistent since this component measures the degree of functional innovation and given that the *"free alternative" variable* measures the ability of software to reproduce the functions of another is the opposite. The negative correlation of the " *package* " variable with axis 2 is explained by the fact that this axis measures the degree of functional reproduction of an existing product. However, the " *package* " variable measures the capacity of software, thanks to the original combination of existing free software, to offer a new use.

**Figure 3: Principal component analysis**

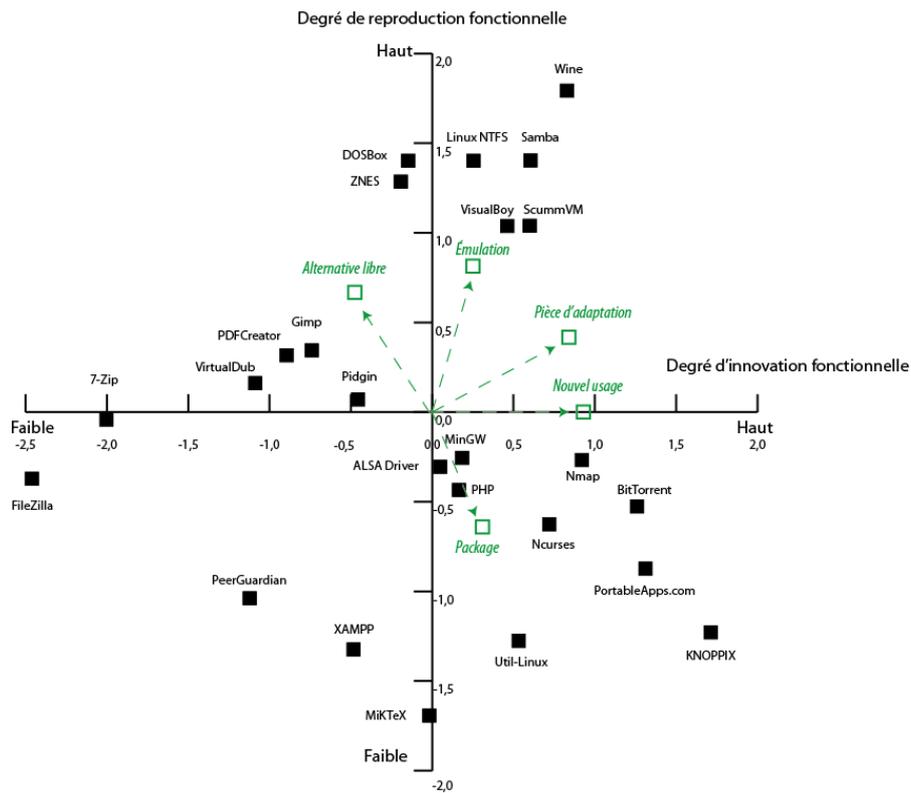

**Table 8: Composition of PCA axes**

| Variables | Components | |
|---|---|---|
| | 1 (degree of functional innovation) | 2 (degree of functional reproduction) |
| New use oriented (variable) | .930 | |
| Adapter piece (varies) | .841 | .418 |
| Emulation (variable) | .251 | .814 |
| Free alternative (varies) | -.476 | .668 |
| Package (varies) | .310 | -.641 |

*Step 3: Comparison of qualitative and quantitative classifications*

- *Goals*

In this step, the objective is to test to what extent the qualitative classification (self-assessment procedure described in step 1) and quantitative (classification based on expert assessments) converge or oppose each other.

- *Data and analytics*

To carry out this step, it is first necessary to assign a qualitative variable (the type of functional innovation) using quantitative variables (five variables previously presented). For this type of procedure, discriminant analysis is a relevant technique. (Droesbeke et al. 2005, p. p. 137). Then it is a question of comparing the qualitative classifications (step 2) and quantitative ones (calculated thanks to the discriminant analysis on the evaluations of the experts).

The discriminant analysis consists of three steps: a test of equality of means, a validation of the study and an analysis of the results of the classification (Burns and Burns 2008, p.: chapitre 25).

First, the test for equality of means ( Wilks ' Lambda ) is significant for all variables (p < 0.05) (see Table 9).

**Table 9: test for equality of means**

|  | Wilks ' Lambda | F | df1 | df2 | Sig . |
|---|---|---|---|---|---|
| Adaptation piece (variable) | .592 | 3,451 | 4 | 20 | .027 |
| Free alternative (variable) | .370 | 8.510 | 4 | 20 | .000 |
| Emulation (varies) | .318 | 10.708 | 4 | 20 | . 000 |
| New use oriented (variable) | .615 | 3.125 | 4 | 20 | .038 |
| Package (varies) | .175 | 23.572 | 4 | 20 | .000 |

Then, the discriminant analysis was relevant since the Box test was also significant (p < 0.05) (see Table 10).

**Table 10: Box test**

| Box's M |  | 61,774 |
|---|---|---|
| F | Approx . | 1,681 |
|  | df1 | 24 |
|  | df2 | 1104.425 |
|  | Sig . | .021 |

- *Results*

Finally, the classification results show that 88% of the software (22 out of 25) were correctly categorized. This confirms that the initial categorization is relevant ( **Table 11**).

**Table 11: results of statistical classification**

| | | Category | Qualitative group | | | | | Total |
|---|---|---|---|---|---|---|---|---|
| | | | Free alternatives | New use oriented | Adapter piece | package | emulator | |
| Original | Number | Free alternative | 5 | 0 | 0 | 0 | 0 | 5 |
| | | New use oriented | 1 | 4 | 0 | 0 | 0 | 5 |
| | | Adapter piece | 0 | 0 | 4 | 1 | 0 | 5 |
| | | package | 0 | 0 | 0 | 5 | 0 | 5 |
| | | emulator | 0 | 1 | 0 | 0 | 4 | 5 |
| | % | Free alternatives | 100.0 | .0 | .0 | .0 | .0 | 100.0 |
| | | New use oriented | 20.0 | 80.0 | .0 | .0 | .0 | 100.0 |
| | | Adapter piece | .0 | .0 | 80.0 | 20.0 | .0 | 100.0 |
| | | package | .0 | .0 | .0 | 100.0 | .0 | 100.0 |
| | | Emulator | .0 | 20.0 | .0 | .0 | 80.0 | 100.0 |
| *Cross-sectional validation* [a] | Number | Free alternative | 5 | 0 | 0 | 0 | 0 | 5 |
| | | New use oriented | 1 | 4 | 0 | 0 | 0 | 5 |
| | | Adapter piece | 1 | 1 | 1 | 1 | 1 | 5 |
| | | package | 0 | 0 | 0 | 5 | 0 | 5 |
| | | emulator | 0 | 1 | 1 | 0 | 3 | 5 |
| | % | Free alternative | 100.0 | .0 | .0 | .0 | .0 | 100.0 |
| | | New use oriented | 20.0 | 80.0 | .0 | .0 | .0 | 100.0 |
| | | Adapter piece | 20.0 | 20.0 | 20.0 | 20.0 | 20.0 | 100.0 |
| | | package | .0 | .0 | .0 | 100.0 | .0 | 100.0 |
| | | emulator | .0 | 20.0 | 20.0 | .0 | 60.0 | 100.0 |

has. *Cross* validation was only carried out for these cases in the analysis. In this validation, each case is classified by the functions derived from each case except this case.
b. 88.0% of cases originally grouped together are correctly classified.
vs. 72.0% of the cases originally grouped (transversal validation) together are correctly classified.

However, the discriminant analysis reveals differences between the qualitative classification and that based on the evaluations of the experts: three software out of 25 seem misclassified (see Table 12).

**Table 12: qualitative and statistical classification**

| Software | Qualitative category | Calculated category |
|---|---|---|
| 7Zip | Free alternative | Free alternatives |
| ALSADriver | Adapter piece | Adapter piece |
| BitTorrent | New use oriented | New use oriented |
| DOSBox | emulator | emulator |
| FileZilla | Free alternative | Free alternative |
| Gimp | Free alternative | Free alternative |
| KNOPPIX | package | package |
| LinuxNTFS | Adapter piece | Adaptation piece |
| MiKTeX | package | package |
| MinGW | Adaptation piece | package |
| Ncurses | emulator | New use oriented |
| Nmap | New use oriented | New use oriented |
| PDFCreator | Free alternative | Free alternative |
| Peer Guardian | New use oriented | New use oriented |
| PHP | New use oriented | New use oriented |
| Pidgin | New use oriented | Free alternative |
| PortableApps | package | package |
| Samba | Adapter piece | Adapter piece |
| ScummVM | Emulator | Emulator |
| utillinux | package | package |

| VirtualDub | Free alternatives | Free alternatives |
|---|---|---|
| VisualBoyAdvance | emulator | emulator |
| Wine | Adapter piece | Adapter piece |
| XAMPP | package | package |
| Zones | emulator | emulator |

First, MinGW was classified as a package ( MinGW brings together GNU tools) while it was placed in the category of adaptation parts: it is also an adaptation of GNU tools for Windows systems. MinGW stands for: Minimalist GNU for Windows (MinGW Team 2008). Then, Ncurses was considered as new use oriented while it was seen as an emulator based on the definition of this library (Free Software Foundation 2007). Finally, Pidgin was classified as a free alternative even though it was among new-use oriented software. Indeed, according to one of the experts *"Pidgin [is] innovative in the sense that it is an entire protocol in a single software"* (Expert 2). Pidgin was probably seen as a free alternative to existing instant messaging clients. In short, the differences between qualitative and statistical classification come from one main reason: there are free software falling into two categories. Nevertheless, the qualitative classification is a good approximation of the average expert judgment.

*Step 4: modeling the types of functional innovation*

- *Goals*

In the previous step, the discriminant analysis made it possible to assign qualitative variables on the basis of quantitative data (aggregate evaluations). It is now a question of modeling the types of functional innovation previously identified (qualitatively) and validated (quantitatively) so that they can be generalized and be reusable for theory and practice.

- *Data*

For each group, the average of the ratings of all the software in the class was calculated. The table belowsummarizes this data.

**Table 13: group characteristics (abbreviated "G")**

|  | Adapter piece (varies) | Free alternative (varies) | Emulation (variable) | New use oriented (variable) | Package (variable) |
|---|---|---|---|---|---|
| **G1: Free alternative** | 56.89 | 74.65 | 17.46 | 41.06 | 26.27 |
| **G2: Adaptation piece** | 92.18 | 70.49 | 60.22 | 61.44 | 28.95 |
| **G3: New use** | 74.04 | 37.71 | 8.61 | 68.61 | 20.82 |
| **G4: Emulator** | 79.55 | 54.29 | 87.12 | 55.15 | 13.53 |
| **G5: Package** | 73.43 | 42.10 | 23.62 | 62.06 | 84.49 |
| **Average** | 73.56 | 55.53 | 35.16 | 57.13 | 37.54 |
| **standard deviation** | 17.61 | 18.12 | 30.21 | 15.38 | 27.48 |

- *Analyzes*

Thanks to the statistics calculated on the evaluations of experts, types of innovation in the field of open source are proposed. Each type was modeled based on the characteristics of the software groups calculated through the aggregation of expert ratings (step 2). For example, for the first group of software (G1), a large majority of experts (nearly 75%) considered that the only added value of this software was the fact of being free alternatives to existing products. These software have a very low, low or average score on all other variables. Conversely, for the third group of software (G3), we see that this software received a high score for the problem-solving and new-use creation variables. However, the latter received a low score on the other dimensions. The table belowsummarizes the characteristics of each type of functional innovation based on aggregated data (see above).

**Table 14: summary of types of functional innovation**

|  | Adapter piece (variable) | Free alternative (varies) | Emulation (varies) | New use (varies) | Package (varies) |
|---|---|---|---|---|---|
| G1: Free alternative (type) | AVERAGE | High | Very low | AVERAGE | Down |
| G2: Adapter piece (type) | Very high | High | High | High | Down |
| G3: New use oriented (type) | High | Down | Very low | High | Down |
| G4: Emulator (type) | High | AVERAGE | Very high | AVERAGE | Very low |
| G5: Package (type) | High | AVERAGE | Down | High | Very high |

| If the value was: |  |
|---|---|
| Less than 20* | Very low |
| Between 20 and 40 | Down |
| Between 40 and 60 | AVERAGE |
| Between 60 and 80 | High |
| Between 80 and 100 | Very high |

**\*Percentage of experts.**

In order to facilitate the appropriation and reuse of this typology, each type of functional innovation should be detailed. First, the free alternative (1) displays a low level on the variables linked to emulation and the combination of products, a medium level on the metrics associated with the resolution of technical problems and with use since these software simply replicates existing software to provide an alternative. The adaptation part (2) presents a very high level for the variable measuring the ability to solve an unsolved problem, a high level on the variables measuring the alternative (this type of software constitutes an alternative to a closed solution), emulation (most of the time this software simulates the operation of a set of software or hardware), new use (this software makes it possible to perform one or more new functions). Conversely, the new use oriented (3) creates a new use by solving a technical problem but does not constitute a free alternative (since no similar functionality existed), nor an emulation or a grouping of products (the values associated with these variables are low or even very low). The emulator (4) offers functionality that existed but in a different form. We note a very high level for the variable relating to emulation and high for the resolution of technical problems. Finally, package (5) presents a very high level for the variable linked to the combination of products, a high level for the variables corresponding to new use and problem solving.

This typology of functional innovation makes it possible to rationally classify open source software on the basis of empirically validated criteria. Figure **Figure 4of** innovation (the adaptation piece).

**Figure 4: synthetic representation of the adaptation part**

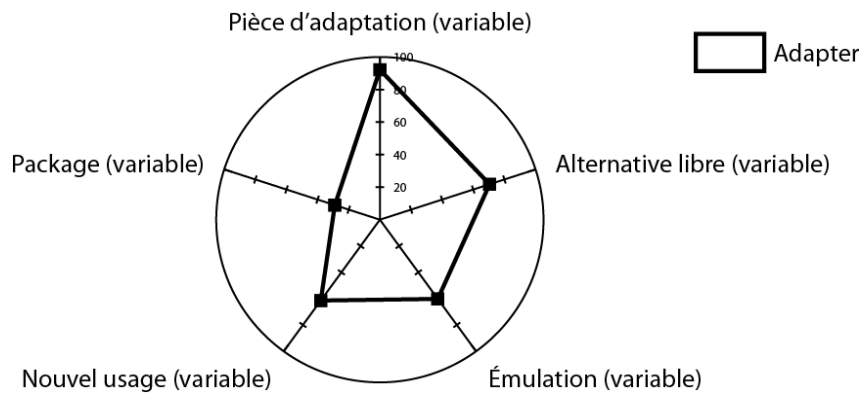

## RESEARCH CONTRIBUTIONS AND IMPLICATIONS

The three main limitations of the literature have been detailed in terms of assessing the innovation of open source software: the apparent confusion between process and innovative product, the lack of criteria for evaluating software innovation and the weak robustness of the evaluation procedures. The first limit was exceeded by the very *design* of this research, which led to a clear delimitation of the field of study: namely the assessment of the product innovation of open source software. Implicitly, this research excludes all elements pertaining to development processes. The other two limitations (lack of criteria, robustness of the procedure) receive a detailed response below.

### Contribution 1: a typology of functional innovation in open source

This article contributes to the growing corpus of research on the specificities of innovation in the software sector. Previous research offered mixed or even completely contradictory results, particularly because the latter were based on a different definition of the notion of innovation. This is due to the inadequacy of traditional economic and management metrics to characterize innovation in the software sector. For example, the incremental/radical diptych (Dewar and Dutton 1986)has serious limitations for analyzing software innovation. In addition to the dualistic nature of the analysis, software will be considered to constitute a *radical innovation* if and only if it is based on a new technology (algorithm, language, model). However, innovation in the field of IT is not limited to technology (Carr 2003)and in the field of software, it can be purely functional (Lippoldt and Stryszowski 2009). Similarly, the analysis of innovation through the maintaining/breaking pair (Christensen 2006)or even the use of the criteria of the Oslo Manual (OECD/Eurostat 2005)only makes sense when there is a market for the analyzed product. When the product is distributed for free, as is the case with most open source software, these models are ineffective.

Furthermore, one of the main limits of dichotomous typologies is the fact that they provide a simplistic (Garcia and Calantone 2002)or binary vision of reality, whereas the latter is complex and multiple (Mingers 2000; Mingers and Brocklesby 1997). For example, a product cannot be *both* a sustaining and disruptive innovation (Christensen 2006; Christensen and Raynor 2003), incremental and radical (Dewar and Dutton 1986; Ettlie et al. 1984)or even continuous and discontinuous (Veryzer Jr. 1998). Conversely, a typology of innovation must allow the description of hybrid forms combining several classes of innovation. For example, MinGW (a software analyzed in this study) can be attached both to the category of *packages* (it is an assembly of several software) and to that of *adaptation parts* (it is software that solves a problem technical).

The evaluation of innovation in the field of software requires the construction of *ad hoc metrics* . This measurement must take into consideration the uses that the technologies allow and not only the technology (Chandy and Tellis 1998)or the relationship to the market (Christensen 2006; OECD/Eurostat 2005). Therefore, this research proposes a new way of approaching the question of software innovation by introducing the concept of *functional added value* ( VAF ). This concept refers to the added value offered by software compared to existing products. To highlight the VAF of a software, it is necessary to list the competing products and to seek in what the product differs or not from the latter from the functional point of view. This concept clarifies the notion of functional innovation proposed in the literature (Lippoldt and Stryszowski 2009).

Previous research has shown that the evaluation of free software requires specific metrics (Capra et al. 2011; Chen et al. 2010; Lundell et al. 2010; Taibi et al. 2007)and more particularly in terms of analysis of innovation (Rossi 2009). In this study, a typology based on the notion of VAF was proposed. It is dedicated to open source but can be extended to other IT fields in subsequent research. This modeling offers the first typology of functional innovation in open source software based on the analysis of numerous products and with the assistance of more than a hundred experts. With current knowledge, it is not reasonably possible to measure the innovation of heterogeneous products such as software, a coffee machine or a handbag with the same metrics. Generic typologies based on notions (eg technological radicalism, patents, trademarks) widely debated or questioned by theory and practice (Garcia and Calantone 2002; Le Masson et al. 2010). This article paves the way for a specialized approach (specific to each sector of activity) to measuring innovation.

### Contribution 2: a framework for innovation modeling (CMI)

The literature has shown that IT evaluation is a complex task (Irani and Love 2001)and constitutes a contemporary challenge (Benlian 2011). Similarly, assessing innovation is a complex activity (Garcia and Calantone 2002). Previous studies on free software innovation presented obvious methodological limitations: whether with regard to the use of the self-evaluation technique itself based on obscure criteria (Deek (Deek and McHugh 2007)or even not empirically validated (Klincewicz 2005)or even a Delphi evaluation based on a very limited number of experts (Rossi 2009).

Conversely, this article adopts a methodological stance combining *hard* and *soft methods* (Mingers 2000)and demonstrates that it is particularly relevant for evaluating innovation. To do this, this study combines two major evaluation techniques often used separately in the literature, namely: self-assessment (Benlian 2011; Klincewicz 2005)and expert judgment (Rossi 2009). Based on the progress of this research, a Framework for Innovation Modeling (CMI) is proposed. CMI includes the following steps: (1) identify and select innovative products, (2) discover and code innovation types, (3) test innovation types, (4) model innovation. In order to ensure appropriation and adaptation of the CMI, it is appropriate to describe the essential elements of each step.

*Identification and selection of innovative products*

The identification of products that seem innovative can be based on: one or more existing product lists listing products known for their performance, a recommendation by experts, one or more professional press articles comparing products. This involves using the principles of theoretical sampling and multiple case study (Eisenhardt 1989; Eisenhardt and Graebner 2007)in order to select products that maximize replication phenomena (repetition of the same type of innovation ), extension (complementarity of products to account for the plurality of innovation in the field studied) and conversely (distinction of products with added value from those that do not, opposition of types of products identified). The objective is to go from a complex and multiple reality to a limited number of cases allowing an in-depth study of each case.

### Discovery and coding of innovation types

Using the data collected during the previous phase, the data will be analyzed manually or using a qualitative analysis tool such as textual analysis software for example. The objective being to discover categories of innovation and derive variables from them while remaining very close to the data, following the principles of rooted theory ( Glaser and Strauss 1967; Miles and Huberman 1994)and multiple case study (Eisenhardt and Graebner 2007). This process should preferably be carried out with the assistance of practitioners or experts in the field of reference.

### Testing types of innovation

The aim here is to derive hypotheses to be validated (or invalidated) and to construct a questionnaire which will be administered to experts in the field. The literature recommends a minimum of 10 experts for Delphi type analyzes (Gallego et al. 2008; Okoli and Pawlowski 2004). These experts may be practitioners or users. The questionnaire should propose a list of products previously selected and classified qualitatively (see step 1). More precisely, it will be necessary to assign a type *a priori* and code it in the form of a nominal variable in a statistical software. Following this, the experts will have to position the products in the relevant category or categories using a Likert scale, for example. It is also possible to use a Boolean variable, however the statistical tests will have to be adapted accordingly. The objective is to obtain an expert consensus. The data can be analyzed using an analysis of variance (or another test depending on the type of variable chosen). The hypotheses will be validated or not. In the event of non-validation, it will be necessary to review the classification and return to step 2. A principal component analysis (or a similar analysis) may be used because it will allow a graphical visualization of the sets of products. The discriminant analysis (DA) will essentially be used to compare the a priori classification and the classification based on the aggregation of expert judgment. The AD will validate or invalidate the classification a priori. The greater the number of correctly classified cases, the more relevant the qualitative classification. Misclassified cases (if any) should be thoroughly analyzed to understand the reasons for misclassification.

### Modeling the types of innovation

From the data collected on the products tested, it will be necessary to model the types of innovation. This modeling should be based on the characteristics of the groups of products resulting from the quantitative classification (assessment by aggregated experts). It can also be based on examples of products so that it is less abstract. The table below summarizes the CMI and its stages.

**Table 15: CMI steps**

| Steps | Detail | Relevant methodologies |
|---|---|---|
| 1. Identify and select innovative products | Identify products that seem innovative. Then select products that maximize replications, extensions or differences. | Theoretical sampling and multiple case study (Eisenhardt 1989; Eisenhardt and Graebner 2007; Fitzgerald 1997) |
| 2. Discover and code types of innovation | Analyze the data (manually or through a dedicated tool) to discover a priori categories and derive variables. | Grounded theory (Glaser and Strauss 1967; Miles and Huberman 1994) |
| 3. Test the types of innovation | Create a questionnaire to test innovation types through expert or user evaluation. Comparison of qualitative and quantitative classifications. | Delphi type evaluation (Rowe and Wright 1999), user evaluation (Beaudry and Pinsonneault 2005); descriptive statistics. |
| 4. Modeling innovation | Model the types of innovation based on statistically aggregated data. | Statistics (Burns and Burns 2008) |

This article has shown that the traditional metrics of innovation economics and management do not make it possible to account for innovation in the field of software. Indeed, in sectors such as IT or traditional activities (Alcaide-Marzal and Tortajada-Esparza 2007), practitioners and academics must adopt an original and innovative posture. The CMI proposes a generic framework for the creation of typologies of innovation contingent on the sector of activity studied. However, more research is needed on the use of multi-methodology for innovation evaluation in IS/IT and other research fields.

**CONCLUSION**

Over the past ten years, users have shown that they are capable of creating and maintaining technological products autonomously (Von Hippel 2005). Wikipedia and Linux are the most popular examples. Through the combination of qualitative and quantitative methodologies, this study contributes to the literature on typology and approach to the evaluation of innovation. However, it has significant limitations that constitute as many *challenges* for research.

The first limitation of this study concerns the type of products analyzed. Indeed, the use of data from *forges* or consumer sites has introduced a bias relating to the type of open source software. This typology relied on open source software created by users. More empirical research is needed to compare software designed by other types of organizations involved in open source such as business networks (Feller et al. 2008)or hybrid communities (Capra et al. 2011). This will make it possible to improve and enrich this typology in order to better reflect the diversity of this field. In addition, this research proves that multimethodology is a relevant approach to assess product innovation. The exercise proved particularly difficult. The proposed framework opens a first way in the use of multimethodology for the evaluation of innovation. However, more research is needed using multimethodology for the evaluation of innovation in IS/IT but also in other fields of research in order to better understand this approach and reveal its potential.

**THANKS**

The author would like to thank Anthony Di Benedetto, Mathias Béjean, Michel Bigand, Laurent Clévy, Marc-Aurèle Darche, Albert David, Serge Druais, Sophie Gautier, Viet Ha, Armand Hatchuel, Ola Henfridsson, Youness Lemrabet, Grégory Lopez, Frank Piller, Frantz Rowe, François-Xavier de Vaujany, Besoa Rabenasolo, Donald White, Stuart Williams and all participants in this study.

Many thanks are extended to the anonymous *French Journal of Information Systems reviewers* whose work allowed this manuscript to evolve radically. Many thanks to Yves Pigneur for his teaching during the revision process.

## Appendix 1: expert profiles

**Table 16: Sectors of activity**

| Fields | Number [1] | Percentage |
|---|---|---|
| Industry and IT services | 62 | 59.62% |
| Universities/Schools | 22 | 21.15% |
| Services (excluding IT) | 11 | 10.58% |
| Research centers | 9 | 8.65% |
| Total | 104 | 100% |

[1] Data was not available for 21 respondents

**Table 17: Positions occupied**

| Jobs | Number [2] | Percentage |
|---|---|---|
| Developers | 29 | 25.22% |
| System administrators | 19 | 16.52% |
| Students | 18 | 15.65% |
| Expert/Architect/Software Engineer | 18 | 15.65% |
| Others | 11 | 9.57% |
| Researchers (academic) | 8 | 6.96% |
| Technical and R&D Directors | 7 | 6.09% |
| Project Managers | 3 | 2.61% |
| Leaders | 2 | 1.74% |
| Total | 115 | 100.00% |

[2] Data was not available for 10 respondents

**Table 18: Development experience**

| Experience | Number | Percentage |
|---|---|---|
| More than 10 years | 41 | 32.80% |
| Between 5 and 10 years | 23 | 18.40% |
| Between 3 and 5 years | 14 | 11.20% |
| Between 1 and 3 years | 13 | 10.40% |
| Less than a year | 5 | 4.00% |
| Do not develop | 29 | 23.20% |
| Total | 125 | 100.00% |

**Table 19: Geographic origin**

| Geographical areas | Number [3] | Percentage |
|---|---|---|
| Europe | 74 | 62.18% |
| Asia | 14 | 11.76% |
| America | 26 | 21.85% |
| Africa | 1 | 0.84% |
| Oceania | 4 | 3.36% |
| Total | 119 | 100.00% |

[3] Data was not available for 6 respondents

## Appendix 2: questions asked

- For free alternatives: This free software is above all a free alternative to existing "proprietary" ( *closed source) software . Example: aMSN is a free clone of the MSN Messenger messaging client.*
- For emulators: *This free software emulates the operation of physical hardware or a set of software applications (eg: operating system, engine, etc.): Example: PCSX2 is software that emulates the operation of the PlayStation 2 console.*
- For packages: *This free software is mainly a collection of free software. Example: EasyPHP combines Apache, MySQL and PHP in a single software.*
- For adaptation parts: *This free software has solved a technical problem that until now had not been solved. Example: Mono is the first software that made it possible to run .Net applications on a Linux system.*
- For those oriented towards new uses: *This free software has created a new use. Example: Mute is P2P (Peer to Peer) software that preserves the anonymity of its users. Mute added the notion of anonymity if we compare it to existing P2P software when it was initiated.*

## Appendix 3: summary of the experts' contributions to the questionnaire

| ID | Expert | Noticed |
|---|---|---|
| 1 | Open source Architect, Thales | Problem of not knowing certain software. |
| 2 | Open source Architect, Thales | Typo in the name of a software: 2SNES instead of ZNES . |
| 3 | Open source Architect, Thales | Citation of Freecode as a source for popular software. |
| 4 | Open source Architect, Thales | Need to have a better balance between "general public" software and "professional" software. |
| 5 | IT researcher | Added "I don't know" option |
| 6 | Open source developer 1 | Correction of mistakes in the English version. |
| 7 | Open source expert, SSLL[16] | Proposal to address developers as a priority (impact on the choice of contacts to whom the questionnaire was sent). |
| 8 | Development Engineer, SSLL | Suggestion of posting the questionnaire online for downloading by participants. |
| 9 | Development Engineer, SSLL | Proposal for the creation of an online form to automate the processing of questionnaires. |
| 10 | Development Engineer, SSLL | Problem of not knowing certain software. |
| 11 | Development Engineer, SSLL | Redundancy of questions 3 and 6 (question amended accordingly). |
| 12 | Development Engineer, SSLL | Presentation problem. |
| 13 | Development Engineer, SSLL | Write a preamble to the questionnaire. |
| 14 | Open source developer 2 | Proposal to translate the questionnaire into Spanish (unfinished). |
| 15 | Member of a forum | Correction of the English version questionnaire. |
| 16 | Researcher, Alcatel-Lucent | Note on the chosen software list. |
| 17 | Researcher, Alcatel-Lucent | Proposal to use an online questionnaire system. |
| 18 | Researcher, Alcatel-Lucent | Addition of a comment box. |
| 19 | Researcher, Alcatel-Lucent | Discussion on the status of technical innovations. |
| 20 | Researcher, Alcatel-Lucent | Modulation of the response depending on the respondent. |
| 21 | Researcher, Alcatel-Lucent | Introduction of the respondent knowledge problem. |
| 22 | Researcher, Alcatel-Lucent | Ambiguity of question 4 on emulation. |
| 23 | Researcher, Alcatel-Lucent | Note on the notion of need. |
| 24 | Researcher, Alcatel-Lucent | Introduction of "I don't know". |
| 25 | Researcher, Alcatel-Lucent | Problem understanding the question about packages. |
| 26 | Researcher, Alcatel-Lucent | Suggestion to have an example for how to answer each question. |
| 27 | Researcher, Alcatel-Lucent | Software history knowledge problem. |
| 28 | Researcher, Alcatel-Lucent | Question 10 deemed too broad. |
| 29 | Researcher, Alcatel-Lucent | Suggestion of new items on the volume of the contribution. |
| 30 | Researcher, Alcatel-Lucent | Suggestion for a new question: professional activity or hobby. |
| 31 | Researcher, Alcatel-Lucent | Comprehension question on the status of the technical problem resolution. |
| 32 | Researcher, Alcatel-Lucent | Question about the status of software porting. |
| 33 | Researcher, Alcatel-Lucent | Question for understanding on the status of emulations. |
| 34 | Researcher, Alcatel-Lucent | Note on the link between innovation and need. |
| 35 | Researcher, Alcatel-Lucent | Proposed rephrasing of the question on packages. |
| 36 | Researcher, Alcatel-Lucent | Confirmation of the proposal to add examples per question. |
| 37 | Researcher, Alcatel-Lucent | General note to further close questions. |
| 38 | Researcher, Alcatel-Lucent | Reformulation of the question on the Top 5 most innovative software. |
| 39 | Researcher, Alcatel-Lucent | Note on emulation (continued). |

---

[16]Services company specializing in Free Software.

## Appendix 4: Aggregated data

| Software | Qualitative category | Adaptation piece (variable) | Alternate (varies) | Emulation (variable) | New Use (variable) | Package (variable) |
|---|---|---|---|---|---|---|
| 7Zip | Free Alternative | 38.6 | 79.2 | 17.4 | 33.8 | 24.7 |
| ALSADriver | Adapt | 78.7 | 42.3 | 30.2 | 49.2 | 39.7 |
| BitTorrent | New use oriented | 86.8 | 28.2 | 8.6 | 84.6 | 13.1 |
| DOSBox | emulator | 81.5 | 69.0 | 82.7 | 48.8 | 18.6 |
| FileZilla | Free alternative | 28.0 | 70.4 | 11.6 | 25.0 | 18.8 |
| Gimp | Free alternative | 75.0 | 80.0 | 11.5 | 47.6 | 25.8 |
| KNOPPIX | Package | 82.6 | 36.4 | 25.6 | 89.3 | 92.8 |
| LinuxNTFS | Adapter | 95.7 | 81.3 | 57.0 | 59.2 | 21.7 |
| MiKTeX | Package | 55.2 | 27.6 | 12.5 | 55.2 | 74.4 |
| MinGW | Adapter | 75.9 | 60.0 | 49.1 | 53.1 | 82.5 |
| Ncurses | Émulateur | 78.4 | 31.4 | 12.1 | 72.5 | 22.4 |
| Nmap | New use oriented | 90.9 | 43.3 | 9.6 | 74.5 | 24.2 |
| PDFCreator | Free alternative | 61.8 | 77.2 | 21.2 | 42.6 | 33.3 |
| Peer Guardian | New use oriented | 46.7 | 38.9 | 2.4 | 40.9 | 25.0 |
| PHP | New use oriented | 67.4 | 46.7 | 10.3 | 70.5 | 19.4 |
| pidgin | New use oriented | 68.6 | 66.7 | 24.4 | 53.2 | 32.8 |
| PortableApps | package | 85.7 | 47.2 | 30.2 | 77.5 | 93.5 |
| Samba | Adapt | 98.1 | 80.0 | 62.0 | 66.7 | 26.9 |
| ScummVM | emulator | 85.3 | 51.4 | 81.1 | 63.2 | 18.2 |
| Utililinux | package | 87.0 | 36.7 | 10.0 | 50.0 | 87.8 |
| VirtualDub | Free alternative | 69.4 | 74.4 | 18.8 | 44.1 | 22.2 |
| VisualBoyAdvance | emulator | 80.0 | 42.3 | 91.8 | 58.6 | 11.4 |
| Wine | Adapt | 96.3 | 78.4 | 91.7 | 70.7 | 27.4 |
| XAMPP | package | 54.3 | 44.7 | 14.3 | 47.4 | 76.1 |
| ZNES | emulator | 71.4 | 54.5 | 92.9 | 50.0 | 5.9 |
| **Average** | | 74 | 56 | 35 | 57 | 38 |
| **standard deviation** | | 17.6 | 18.1 | 30.2 | 15.4 | 27.5 |
| **Maximum** | | 98.1 | 81.3 | 92.9 | 89.3 | 93.5 |
| **Minimum** | | 28.0 | 27.6 | 2.4 | 25.0 | 5.9 |

### Annex 5: ANOVA of categories

| | | sum of squares | df | Mean of squares | F | Sig . |
|---|---|---|---|---|---|---|
| Adapter piece (varies) | Between groups | 3164.170 | 4 | 791.043 | 3.451 | .027 |
| | In groups | 4584.469 | 20 | 229.223 | | |
| | Total | 7748.640 | 24 | | | |
| Free alternative (varies) | Between groups | 5168.638 | 4 | 1292.159 | 8.510 | .000 |
| | In groups | 3036.874 | 20 | 151.844 | | |
| | Total | 8205.511 | 24 | | | |
| Emulation (variable) | Between groups | 15549.727 | 4 | 3887.432 | 10.708 | .000 |
| | In groups | 7260.716 | 20 | 363.036 | | |
| | Total | 22810.443 | 24 | | | |
| New use (varies) | Between groups | 2275.034 | 4 | 568.759 | 3.125 | .038 |
| | In groups | 3640.026 | 20 | 182.001 | | |
| | Total | 5915.060 | 24 | | | |
| Package (varies) | Between groups | 15571.418 | 4 | 3892.855 | 23.572 | .000 |
| | In groups | 3303.005 | 20 | 165.150 | | |
| | Total | 18874.423 | 24 | | | |

### Appendix 6: eigenvalues of principal component analysis

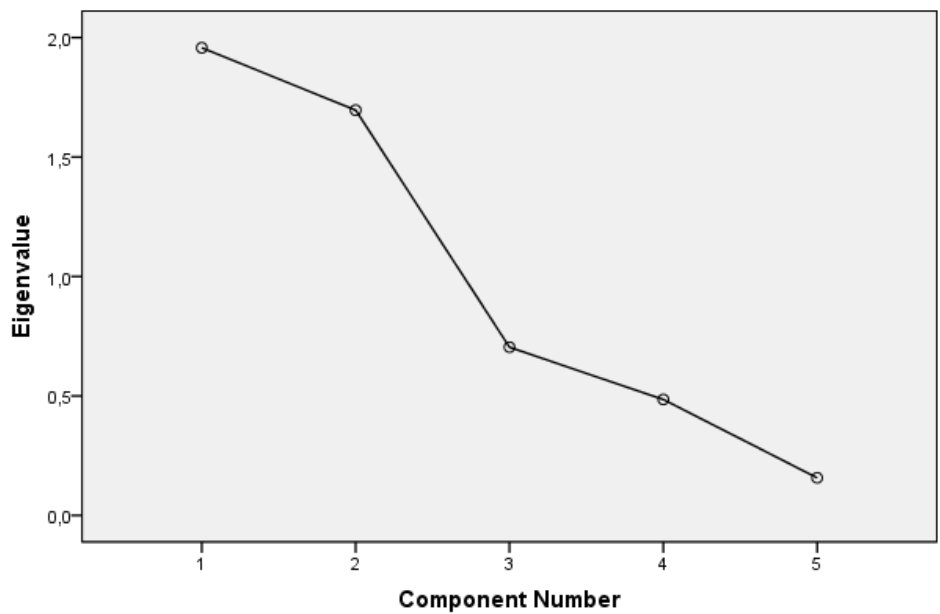

Afuah, A. 2000. "How Much Do Your Co-Opetitors' Capabilities Matter in the Face of Technological Change?," *Strategic Management Journal* (21:3), pp. 397-404.

Agerfalk, P. J., and Fitzgerald, B. 2008. "Outsourcing to an Unknown Workforce: Exploring Opensourcing as a Global Sourcing Strategy," *Management Information System Quarterly* (32:2), pp. 385-409.

Alcaide-Marzal, J., and Tortajada-Esparza, E. 2007. "Innovation Assessment in Traditional Industries. A Proposal of Aesthetic Innovation Indicators," *Scientometrics* (72:1), pp. 33-57.

Allen, R. C. 1983. "Collective Invention," *Journal of Economic Behavior and Organization*:4), pp. 1-24.

Baldwin, C., and Von Hippel, E. 2011. "Modeling a Paradigm Shift: From Producer Innovation to User and Open Collaborative Innovation," *Organization Science* (22:6), pp. 1399-1417.

Beaudry, A., and Pinsonneault, A. 2005. "Understanding User Responses to Information Technology: A Coping Model of User Adaptation," *MIS Quarterly* (29:3), pp. 493-524.

Benkeltoum, N. 2011a. *Gérer Et Comprendre L'open Source*. Paris: Presses des Mines.

Benkeltoum, N. 2011b. "Regards Sur Les Stratégies De Détournement Dans L'industrie Open Source," *Vie et sciences de l'entreprise*:187), pp. 72-91.

Benlian, A. 2011. "Is Traditional, Open-Source, or on-Demand First Choice[Quest] Developing an Ahp-Based Framework for the Comparison of Different Software Models in Office Suites Selection," *European Journal of Information Systems* (20:5), pp. 542-559.

Bitzer, J., and Schröder, P. J. H. 2005. "Bug-Fixing and Code-Writing: The Private Provision of Open Source Software," *Information Economics and Policy* (17:3), pp. 389-406.

Bogers, M., Afuah, A., and Bastian, B. 2010. "Users as Innovators: A Review, Critique, and Future Research Directions," *Journal of Management* (36:4), pp. 857-875.

Bonaccorsi, A., Giannangeli, S., and Rossi, C. 2006. "Entry Strategies under Competing Standards: Hybrid Business Models in the Open Source Software Industry," *Management Science* (52:7), pp. 1085-1098.

Burns, R., and Burns, R. 2008. *Business Research Methods and Statistics Using Spss*. London: Sage Publications Ltd.

Capiluppi, A., Boldyreff, C., and Stol, K.-J. 2011. "Successful Reuse of Software Components: A Report from the Open Source Perspective," in *Open Source Systems: Grounding Research*, S. Hissam, B. Russo, M. de Mendonça Neto and F. Kon (eds.). Springer Boston, pp. 159-176.

Capra, E., Francalanci, C., Merlo, F., and Rossi-Lamastra, C. 2011. "Firms' Involvement in Open Source Projects: A Trade-Off between Software Structural Quality and Popularity," *Journal of Systems and Software* (84:1), pp. 144-161.

Carr, N. G. 2003. "It Doesn't Matter," *Harvard Business Review* (81:5), pp. 41-49.

Chandy, R. K., and Tellis, G. J. 1998. "Organizing for Radical Product Innovation: The Overlooked Role of Willingness to Cannibalize," *Journal of Marketing Research* (35:November), pp. 474-487.

Chen, D., Shams, S., Carmona-Moreno, C., and Leone, A. 2010. "Assessment of Open Source Gis Software for Water Resources Management in Developing Countries," *Journal of Hydro-environment Research* (4:3), pp. 253-264.

Christensen, C. M. 1997. *The Innovator's Dilemma: When New Technologies Cause Great Firms to Fail*. Boston, Massachusetts: Harvard Business School Press.

Christensen, C. M. 2006. "The Ongoing Process of Building a Theory of Disruption," *Journal of Product Innovation Management* (23), pp. 39-55.

Christensen, C. M., and Overdorf, M. 2000. "Meeting the Challenge of Disruptive Change," *Harvard Business Review* (March–April), pp. 66-76.

Christensen, C. M., and Raynor, M. E. 2003. *The Innovator's Solution: Creating and Sustaining Successful Growth*. Boston, Massachusetts: Harvard Business School Press.

CIGREF. 2011. "Maturité Et Gouvernance De L'open Source - La Vision Des Grandes Entreprises," CIGREF, Paris.

Codenie, W., Pikkarainen, M., Boucart, N., and Deleu, J. 2011. "Introduction," in *The Art of Software Innovation - Eight Practice Areas to Inspire Your Business*, M. Pikkarainen, W. Codenie, N. Boucart and J.A. Heredia Alvaro (eds.). Berlin Heidelberg: Springer, pp. 1-18.

Dahlander, L., and Wallin, M. W. 2006. "A Man on the Inside: Unlocking Communities as Complementary Assets," *Research Policy* (35:7), pp. 1243-1259.

Dahlin, K. B., and Behrens, D. M. 2005. "When Is an Invention Really Radical? Defining and Measuring Technological Radicalness," *Research Policy*:34), pp. 717-737.

Danneels, E. 2004. "Disruptive Technology Reconsidered: A Critique and Research Agenda," *Journal of Product Innovation Management* (21), pp. 246-258.

Danneels, E. 2006. "From the Guest Editor Dialogue on the Effects of Disruptive Technology on Firms and Industries," *Journal of Product Innovation Management* (23), pp. 2-4.

Deek, F. P., and McHugh, J. A. 2007. *Open Source: Technology and Policy*. New York: Cambridge University Press.

Del Bianco, V., Lavazza, L., Morasca, S., and Taibi, D. 2009. "Quality of Open Source Software: The Qualipso Trustworthiness Model," *Open Source Ecosystems: Diverse Communities Interacting, 13th International Conference on Open Source Systems, IFIP*, C. Boldyreff, K. Crowston, B. Lundell and A.I. Wasserman (eds.), Skövde, Sweden: Springer, pp. 199-212.

Dewar, R. D., and Dutton, J. E. 1986. "The Adoption of Radical and Incremental Innovations: An Empirical Analysis," *Management Science* (32:11), pp. 1422-1433.

Droesbeke, J. J., Lejeune, M., and Saporta, G. 2005. *Méthodes Statistiques Pour Données Qualitatives*. Editions Technip.

Ebert, C. 2007. "Open Source Drives Innovation," *IEEE Software* (24:3), pp. 105-109.

Ebert, C. 2008. "Open Source Software in Industry," *IEEE Software* (25:3), pp. 52-53.

Eisenhardt, K. M. 1989. "Building Theories from Case Study Research," *The Academy of Management Review* (14:4), pp. 532-550.

Eisenhardt, K. M., and Graebner, M. E. 2007. "Theory Building from Cases: Opportunities and Challenges," *Academy of Management Journal* (50:1), pp. 25-32.

Ettlie, J. E., Bridges, W. P., and O'Keefe, R. D. 1984. "Organization Strategy and Structural Differences for Radical Versus Incremental Innovation," *Management Science* (30:6), pp. 682-695.

Fearon, C., and Philip, G. 1998. "Self Assessment as a Means of Measuring Strategic and Operational Benefits from Edi: The Development of a Conceptual Framework," *European Journal of Information Systems* (7:1), pp. 5-16.

Feller, J., Finnegan, P., Fitzgerald, B., and Hayes, J. 2008. "From Peer Production to Productization: A Study of Socially Enabled Business Exchanges in Open Source Service Networks," *Information Systems Research* (19:4), pp. 475-493.

Fitz-Gerald, S. 2010. "K. Vadera, B. Gandhi, Open Source Technology (2009) University Science Press, Laxmi Publications, New Delhi, Pp. 173," *International Journal of Information Management* (30:4), p. 374.

Fitzgerald, B. 1997. "The Use of Systems Development Methodologies in Practice: A Field Study," *Information Systems Journal* (7:3), pp. 201-212.

Fitzgerald, B. 2006. "The Transformation of Open Source Software," *Management Information System Quarterly* (30:3), pp. 587-598.

Franke, N., Schreier, M., and Kaiser, U. 2010. "The "I Designed It Myself" Effect in Mass Customization," *Management Science* (56:1), pp. 125-140.

Franke, N., and von Hippel, E. 2003. "Satisfying Heterogeneous User Needs Via Innovation Toolkits: The Case of Apache Security Software," *Research Policy* (32), pp. 1199-1215.

Free Software Foundation. 2007. "Announcing Ncurses 5.6.", from http://www.gnu.org/software/ncurses/ncurses.html

Fuggetta, A. 2003. "Controversy Corner: Open Source Software an Evaluation," *The Journal of Systems and Software*:66), pp. 77-90.

Gallego, M. D., Luna, P., and Bueno, S. 2008. "Designing a Forecasting Analysis to Understand the Diffusion of Open Source Software in the Year 2010," *Technological Forecasting and Social Change* (75:5), pp. 672-686.

Garcia, R., and Calantone, R. 2002. "A Critical Look at Technological Innovation Typology and Innovativeness Terminology: A Literature Review," *Journal of Product Innovation Management* (19), pp. 110-132.

Gary, K., Enquobahrie, A., Ibanez, L., Cheng, P., Yaniv, Z., Cleary, K., Kokoori, S., Muffih, B., and Heidenreich, J. 2011. "Agile Methods for Open Source Safety-Critical Software," *Software: Practice and Experience* (41:9), pp. 945-962.

Glaser, B. G., and Strauss, A. L. 1967. *The Discovery of Grounded Theory: Strategies for Qualitative Research*. Transaction Publishers.

Haefliger, S., von Krogh, G., and Spaeth, S. 2008. "Code Reuse in Open Source Software," *Management Science* (54:1), pp. 180-193

Hamerly, J., Paquin, T., and Walton, S. 1999. "Freeing the Source: The Story of Mozilla," in *Open Sources: Voices from the Open Source Revolution,* C. DiBona, S. Ockman and M. Stone (eds.). Sebastapol, CA: O'Reilly, pp. 197-206.

Hars, A., and Ou, S. 2002. "Working for Free? Motivations for Participating in Open-Source Projects," *International Journal of Electronic Commerce* (6.:3), pp. 25-39.

Heredia Alvaro, J. A., and Pikkarainen, M. 2011. "The Art of Innovation Incubation," in *The Art of Software Innovation - Eight Practice Areas to Inspire Your Business,* M. Pikkarainen, W. Codenie, N. Boucart and J.A. Heredia Alvaro (eds.). Berlin Heidelberg: Springer, pp. 104-122.

Hicks, C., and Pachamanova, D. 2007. "Back-Propagation of User Innovations: The Open Source Compatibility Edge," *Business Horizons* (50), pp. 315–324.

Iivari, N. 2010. "Discursive Construction of 'User Innovations' in the Open Source Software Development Context," *Information and Organization* (20:2), pp. 111-132.

Irani, Z., and Love, P. E. D. 2001. "Information Systems Evaluation: Past, Present and Future," *European Journal of Information Systems* (10:4), pp. 183-188.

Janamanchi, B., Katsamakas, E., Raghupathi, W., and Gao, W. 2009. "The State and Profile of Open Source Software Projects in Health and Medical Informatics," *International Journal of Medical Informatics* (78:7), pp. 457-472.

Jeppesen, L. B., and Frederiksen, L. 2006. "Why Do Users Contribute to Firm-Hosted User Communities? The Case of Computer-Controlled Music Instruments," *Organization Science*:17), pp. 45-63.

Klincewicz, K. 2005. "Innovativeness of Open Source Software Projects." School of Innovation Management, Tokyo Institute of Technology, pp. 11-31.

Kogut, B., and Metiu, A. 2000. "The Emergence of E-Innovation: Insights from Open Source Software Development." Philadelphia: Wharton School, University of Pennsylvania.

Kogut, B., and Metiu, A. 2001. "Open-Source Software Development and Distributed Innovation," *Oxford Review of Economic Policy* (17:2), pp. 248-263.

Kozinets, R. V. 2002. "The Field Behind the Screen: Using Netnography for Marketing Research in Online Communities," *Journal of Marketing Research* (39:1), pp. 61-72.

Kozinets, R. V. 2006. "Click to Connect: Netnography and Tribal Advertising," *Journal of Advertising Research* (46:3), pp. 279-288.

Lakhani, K., and von Hippel, E. 2003. "How Open Source Software Works: "Free" User-to-User Assistance," *Research Policy* (32:6), pp. 923-943.

Le Masson, P., Weil, B., and Hatchuel, A. 2010. *Strategic Management of Innovation and Design*. New York: Cambridge University Press.

Le Texier, T., and Versailles, D. W. 2009. "Open Source Software Governance Serving Technological Agility: The Case of Open Source Software within the Dod," *International Journal of Open Source Software and Processes (IJOSSP)* (1:2), pp. 14-27.

Lee, G. K., and Cole, R. E. 2003. "From a Firm-Based to a Community-Based Model of Knowledge Creation: The Case of the Linux Kernel Development," *Organization Science* (14:6), pp. 633-649.

Lee, M. L., and Davis, J. 2003. "Evolution of Open Source Software: A Study on the Samba Project," *Système d'Information et Management* (8:1), pp. 43-62.

Lee, S.-Y. T., Kim, H.-W., and Gupta, S. 2009. "Measuring Open Source Software Success," *Omega* (37:2), pp. 426-438.

Lindman, J., Rossi, M., and Puustell, A. 2011. "Matching Open Source Software Licenses with Corresponding Business Models," *IEEE Software* (28:4), pp. 31-35.

Lippoldt, D., and Stryszowski, P. 2009. *Innovation in the Software Sector*. OECD Publications.

Lisein, O., Pichault, F., and Desmecht, J. 2009. "Les Business Models Des Sociétés De Services Actives Dans Le Secteur Open Source," *Systèmes d'Information et Management* (14:2), pp. 7-38.

Lundell, B., Lings, B., and Lindqvist, E. 2010. "Open Source in Swedish Companies: Where Are We?," *Information Systems Journal* (20:6), pp. 519-535.

MacCormack, A., Rusnak, J., and Baldwin, C. Y. 2006. "Exploring the Structure of Complex Software Designs: An Empirical Study of Open Source and Proprietary Code," *Management Science* (52:7), pp. 1015-1030.

Midha, V., and Palvia, P. 2012. "Factors Affecting the Success of Open Source Software," *Journal of Systems and Software* (85:4), pp. 895–905.

Miles, M. B., and Huberman, A. M. 1994. *Qualitative Data Analysis*. SAGE Publications, Inc.

Mingers, J. 2000. "Variety Is the Spice of Life: Combining Soft and Hard or/Ms Methods," *International Transactions in Operational Research* (7:6), pp. 673-691.

Mingers, J. 2003. "The Paucity of Multimethod Research: A Review of the Information Systems Literature," *Information Systems Journal* (13:3), pp. 233-249.

Mingers, J., and Brocklesby, J. 1997. "Multimethodology: Towards a Framework for Mixing Methodologies," *Omega* (25:5), pp. 489-509.

MinGW Team. 2008. "Minimalist Gnu for Windows." from http://www.mingw.org/

Miralles, F., Sieber, S., and Valor, J. 2006. "An Exploratory Framework for Assessing Open Source Software Adoption," *Système d'Information et Management* (11:1), pp. 85-103.

Nambisan, S., Agarwal, R., and Tanniru, M. 1999. "Organizational Mechanisms for Enhancing User Innovation in Information Technology," *MIS Quarterly* (23:3), pp. 365-395.

Nuvolari, A. 2004. "Collective Invention During the British Industrial Revolution: The Case of the Cornish Pumping Engine," *Cambridge Journal of Economics* (28:3), pp. 347-363.

OECD/Eurostat, L. 2005. *Oslo Manual: Guidelines for Collecting and Interpreting Innovation Data, 3rd Edition, the Measurement of Scientific and Technological Activities*. OECD Publishing.

Okoli, C., and Pawlowski, S. D. 2004. "The Delphi Method as a Research Tool: An Example, Design Considerations and Applications," *Information & Management* (42:1), pp. 15-29.

Osterloh, M., and Rota, S. 2007. "Open Source Software Development: Just Another Case of Collective Invention?," *Research Policy* (36), pp. 157-171.

Paulson, J. W., Succi, G., and Eberlein, A. 2004. "An Empirical Study of Open-Source and Closed-Source Software Products," *IEEE Trans. Softw. Eng.* (30:4), pp. 246-256.

Pinsonneault, A., and Kraemer, K. L. 1993. "Survey Research Methodology in Management Information Systems: An Assessment," *Journal of Management Information Systems* (10:2), pp. 75-105.

Rietveld, T., and Van Hout, R. 1993. *Statistical Techniques for the Study of Language and Language Behaviour*. De Gruyter.

Rossi, C. 2009. "Software Innovativeness. A Comparison between Proprietary and Free/Open Source Solutions Offered by Italian Smes," *R&D Management* (39:2), pp. 153-169.

Rowe, G., and Wright, G. 1996. "The Impact of Task Characteristics on the Performance of Structured Group Forecasting Techniques," *International Journal of Forecasting* (12:1), pp. 73-89.

Rowe, G., and Wright, G. 1999. "The Delphi Technique as a Forecasting Tool: Issues and Analysis," *International Journal of Forecasting* (15), pp. 353–375.

Shah, S. K. 2006. "Motivation, Governance, and the Viability of Hybrid Forms in Open Source Software Development," *Management Science* (52:7), pp. 1000-1014.

Sigfridsson, A., and Sheehan, A. 2011. "On Qualitative Methodologies and Dispersed Communities: Reflections on the Process of Investigating an Open Source Community," *Information and Software Technology* (53:9), pp. 981-993.

Soens, W. 2011. "The Art of Idea Harvesting," in *The Art of Software Innovation - Eight Practice Areas to Inspire Your Business*, M. Pikkarainen, W. Codenie, N. Boucart and J.A. Heredia Alvaro (eds.). Berlin Heidelberg: Springer, pp. 33-48.

Spaeth, S., Stuermer, M., and von Krogh, G. 2010. " Enabling Knowledge Creation through Outsiders: Towards a Push Model of Open Innovation," *International Journal of Technology Management* (52:3/4), pp. 411-431.

Spiller, D., and Wichmann, T. 2002. "Floss Final Report - Part 3: Basics of Open Source Software Markets and Business Models," Berlecon Research GmbH, pp. 1-58.

Spinellis, D., Gousios, G., Karakoidas, V., Louridas, P., Adams, P. J., Samoladas, I., and Stamelos, I. 2009. "Evaluating the Quality of Open Source Software," *Electronic Notes in Theoretical Computer Science* (233), pp. 5-28.

Stallman, R. 1983. "Free Unix." from http://www.gnu.org/gnu/initial-announcement.html

Stallman, R. 2002. "Obstructing Custom Adaptation of Programs," in *Free Software, Free Society: Selected Essays of Richard M. Stallman,* J. Gay (ed.). Boston: Gnu Press, pp. 124-125.

Stewart, K. J., and Gosain, S. 2006. "The Impact of Ideology on Effectiveness in Open Source Software Development Teams," *MIS Quarterly* (30:2), pp. 291-314.

Stuermer, M. 2009. "How Firms Make Friends: Communities in Private-Collective Innovation, Doctoral Dissertation," in: *Chair of Strategic Management and Innovation, Department of Management, Technology, and Economics*. Zürich: ETH Zürich.

Stuermer, M., Sebastian, S., and von Krogh, G. 2009. "Extending Private-Collective Innovation: A Case Study," *R&D Management* (39:2), pp. 170-191.

Taibi, D., Lavazza, L., and Morasca, S. 2007. "Openbqr: A Framework for the Assessment of Oss," in *Open Source Development, Adoption and Innovation,* IFIP (ed.). Boston: Springer, pp. 173-186.

Thakur, D. 2012. "A Limited Revolution — the Distributional Consequences of Open Source Software in North America," *Technological Forecasting and Social Change* (79:2), pp. 244-251.

The Economist. 2006. "Open-Source Business. Open, but Not as Usual."

Ulhoi, J. P. 2004. "Open Source Development: A Hybrid in Innovation and Management Theory," *Management Decision* (42:9), pp. 1095-1114.

Veryzer Jr., R. W. 1998. "Key Factors Affecting Customer Evaluation of Discontinuous New Products," *Journal of Product Innovation Management* (15:2), pp. 136-150.

Von Hippel, E. 1988. *The Sources of Innovation*. New York: Oxford University Press.

Von Hippel, E. 2002. "Open Source Projects as Horizontal Innovation Networks by and for Users." MIT, pp. 1-26.

Von Hippel, E. 2005. *Democratizing Innovation*. Cambridge: MIT Press.

Von Hippel, E. 2010. "Chapter 9 - Open User Innovation," in *Handbook of the Economics of Innovation,* B.H. Hall and N. Rosenberg (eds.). North-Holland, pp. 411-427.

Von Hippel, E., and Von Krogh, G. 2003. "Open Source Software and the "Private-Collective" Innovation Model: Issues for Organization Science," *Organization Science* (14:2), pp. 209-223.

Von Krogh, G., and Spaeth, S. 2007. "The Open Source Software Phenomenon: Characteristics That Promote Research," *Journal of Strategic Information Systems* (16:3), pp. 236-253.

Von Krogh, G., Spaeth, S., and Lakhani, K. 2003. "Community, Joining, and Specialization in Open Source Software Innovation: A Case Study," *Research Policy* (32:7), pp. 1217-1241.

West, J., and O'Mahony, S. 2005. "Contrasting Community Building in Sponsored and Community Founded, Open Source Projects," *Proceedings of the 38th Annual Hawaii International Conference on System Sciences*, Waikoloa, Hawaii IEEE, pp. 1-10.